\documentclass[prl,twocolumn,floatfix]{revtex4}
\usepackage{graphicx}

\begin{document}

\title{Mesoscopic oscillator in U-shape with giant persistent current}

\author{Y. Z. He and C. G. Bao\footnote{The corresponding author}}

\affiliation{
State Key Laboratory of Optoelectronic Materials and Technologies,
and Department of Physics,
Sun Yat-Sen University,
Guangzhou, 510275, P.R. China}

\begin{abstract}
A mesoscopic oscillator in U-shape has been proposed and studied.
Making use of a magnetic flux together with a potential of
confinement, the electron contained in the oscillator has been
localized initially and an amount of energy has been thereby stored.
Then a sudden cancellation of both the potential and the flux may
cause an initial current which initiates a periodic motion of the
electron from one end of the U-oscillator to the opposite end, and
repeatedly. The period is adjustable. The current associated with
the periodic motion can be tuned very strong (say, more than two
orders larger than the current of the usual Aharonov-Bohm
oscillation). Related theory and numerical results are presented.
\end{abstract}


\maketitle

Due to the great progress in experiments, a few given number of
electrons can be captured and confined in various ingenious
artificial mesoscopic devices (say, quantum dots, wires, rings, and
more complicated coupled dots, Mobius rings,
etc.)~\cite{r_LA98,r_LA2K,r_HP,r_KUF,r_MD,r_FA,r_HEA,r_VS}. These
devices are basic elements in micro-industry. In developing
micro-techniques a crucial point is the counting of time. In this
paper a micro-oscillator in U-shape is proposed, which might work as
a mesoscopic pendulum.

\begin{figure}[tbp]
 \centering
 \resizebox{0.95\columnwidth}{!}{\includegraphics{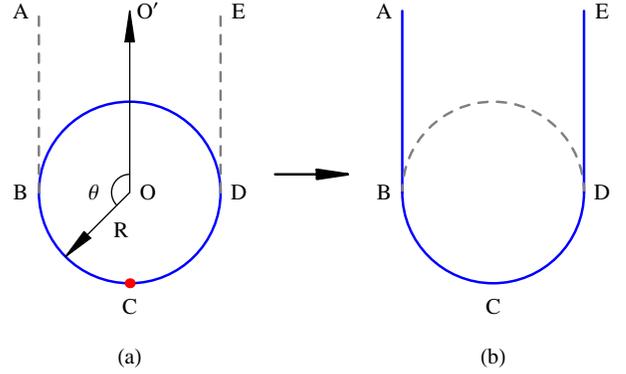}}
 \caption{(Color online.) A sketch of the oscillator. The device in
(a) is in fact a ring, where the two arms (AB and DE in dash line)
are blocked. Whereas in (b) the upper half circle (in dash line) is
blocked and the previous block on the arms is released. Thereby the
ring is transformed to a U-pipe. }
 \label{eufig1}
\end{figure}

It is recalled that, for classical motion, the crucial point of an
oscillator is the storage of an amount of energy which can be
transformed to kinetic energy later. For the classical pendulum, an
amount of potential energy from gravity has been stored in advance,
and will be transformed to the energy of swinging afterward. For a
quantum oscillator, the crucial point is also how to store the
energy. For this purpose, based on the idea suggested in the
ref.~\cite{r_HYZ}, a device is proposed as sketched in
Fig.\ref{eufig1}. The device is a ring together with two arms (AB
and DE), wherein a few free electrons are contained. Previously, the
arms are exactly blocked (say, by electrodes with high voltage), the
ring is threaded by a magnetic flux $\Phi $, and a strong potential
$V$ is applied so that the electron is localized in the bottom of
the ring (close to point C in Fig.1a). In this way, an amount of
energy (which can be quite large) has been stored as shown later.
Suddenly, both $\Phi $ and $V$ are cancelled, and the block on the
arms is released. Instead, the upper half circle of the ring is
blocked. With this sudden change, the ring is transformed to a
U-oscillator (Fig.\ref{eufig1}b). The previously stored energy will
motivate an oscillation of the electrons, as we shall see, from one
end (A) to the other end (E). Dissipation is assumed to be
negligible (namely, a superconducting device). Then, the oscillation
will proceed on in an exact periodic way. Related theories together
with numerical results are as follows.

For simplicity, it is assumed that only one free electron with an
effective mass $m^*$ is contained in the device. Let the radius of
the ring be $R$ and the lengths of both arms be $l$. The ring and
arms are very narrow so that they are quasi-one-dimensional.
Previously, the Hamiltonian defined on the ring associated with
Fig.\ref{eufig1}a reads
\begin{equation}
 H_{init}=G(-i\frac{\partial}{\partial\theta}+\Phi)^2+V(\theta)
 \label{e_Hinit}
\end{equation}
Where $\theta $ is the azimuthal angle of the electron (the vector
$\overrightarrow{OO'}$\ pointing up in Fig.\ref{eufig1}a has $\theta
=0$), $G=\frac{\hbar ^{2}}{ 2m^{\ast }R^{2}}$. $\Phi $ is the
magnetic flux in the unit $\Phi _{0}=hc/e $. $V(\theta )=0$ if $\pi
-d\leq \theta \leq \pi +d$, or $=V_{o}$ otherwise. Where $d$
measures the width of the confinement, $V_{o}$ is a sufficiently
large positive number so that the electron in the ground state $\psi
_{g}$ with energy $E_{g}>0$ is strongly localized previously.

In order to obtain $\psi _{g}$, $H_{init}$ is diagonalized by using
the set $|k\rangle =e^{ik\theta }/\sqrt{2\pi }$ as basis functions,
where $k$\ are integers ranging from $-\infty $ to $+\infty$. The
matrix elements read
\begin{widetext}
\begin{equation}
 \langle k^{\prime }|H_{init}|k\rangle =
 \left\{
 \begin{array}{lll}
 -\frac{V_{o}}{\pi (k-k^{\prime })} \cos ((k-k^{\prime })\pi )\sin
((k-k^{\prime })d) && (\mbox{if } k\neq k')  \\
 G(k+\Phi )^{2}+V_{o}(1-d/\pi ) && (\mbox{if } k=k')
 \end{array}
 \right.
 \label{e_MtxHinit}
\end{equation}
\end{widetext}

For numerical calculation, the range of $k$ must be limited, and it
was found that $-40\leq k\leq 40$ is sufficient to provide accurate
results (say, have at least four effective figures.). After the
diagonalization $ E_{g}$ and $\psi _{g}=\sum_{k}C_{k}^{g}|k\rangle$
can be obtained.

Suddenly, the well is removed and $V(\theta )$ becomes zero
everywhere, the flux $\Phi $ is also removed, and at the same time
the block on the arms is released while the upper-half of the ring
is exactly blocked. This leads to a change of the path of motion.
Accordingly, the Hamiltonian is changed from $H_{init}$ to
\begin{equation}
 H_{evol}=-\frac{\hbar ^{2}}{2m^*}\frac{\partial^2}{\partial s^2}
 \label{e_Hevol}
\end{equation}
defined in the U-pipe, where $s$ is the distance of the electron
apart from A (say, when the electron locates at the lower-half ring,
$s=l+R(\theta -\pi /2)$). After the change $\psi _{g}$ is no more an
eigen-state of the new Hamiltonian, therefore the electron begins to
evolve.

The formal time-dependent solution of the new Hamiltonian with the
initial state $\psi _{g}$ reads
\begin{equation}
 \Psi (s,t)=e^{-iH_{evol}t/\hbar }\psi _{g}
 \label{e_PsiH}
\end{equation}
This formal solution can be rewritten in an applicable form if we
know all the eigen-states of $H_{evol}$. These eigen-states read
simply $|j\rangle = \sqrt{2/L}\sin (p_{j}s)$, where $L\equiv
2l+R\pi$ is the total length of the U-pipe and $s$ is ranged from
zero to $L$, $\ p_{j}=j\pi /L$ and $j$\ is a positive nonzero
integer. With them, Eq.(\ref{e_PsiH}) becomes
\begin{equation}
 \Psi (s,t)=\sum_j e^{-iE_j t/\hbar }|j\rangle \langle j|\psi _g \rangle
 \label{e_PsiE}
\end{equation}
where $E_{j}=\hbar ^{2}p_{j}^{2}/(2m^*)$. The summation of $j$ in
Eq.(\ref{e_PsiE}) is in principle from $1$ to $+\infty $. However,
in numerical calculation $j$ can be confined within a range (say,
from $1$ to $80$).

From Eq.(\ref{e_PsiE}) the time-dependent density
\begin{equation}
 \begin{array}{lll}
 \rho (s,t)
 &\equiv& |\Psi (s,t)|^2 \\
 &=&\frac{2}{L} \sum_{j,j'} \sin(js\pi /L)\sin (j's\pi /L) \\
 & &\mbox{Re}\{e^{-i(j-j')(j+j^{\prime })\tau }
    \langle \psi _g |j'\rangle \langle j|\psi _g \rangle \}
 \label{e_Rho}
 \end{array}
\end{equation}
where only the real part of the right hand side is contributed.
$\tau =t/t_{o}$, and $t_{o}\equiv 2m^{\ast }L^{2}/(\pi ^{2}\hbar )$
is used as a unit of time. The time-dependent current defined from
the conservation of mass reads
\begin{equation}
 \begin{array}{lll}
 J(s,t)
 &=&\frac{4}{\pi } \sum_{j,j'}\cos (js\pi /L)\sin (j's\pi /L)\ j \\
 & &\mbox{Im}\{e^{-i(j-j')(j+j')\tau }
    \langle \psi _g |j'\rangle \langle j|\psi _g \rangle \}
 \end{array}
 \label{e_J}
\end{equation}
where only the imaginary part is contributed. In Eq.(\ref{e_J}) the
unit of current is $t_{o}^{-1}$. From now on, we shall use $\tau$ to
measure the time. Obviously, both $\rho$ and $J$ are strictly
periodic, the period of $\tau$ is $2\pi$.

Since $\psi _g$ is localized in the bottom of the U-pipe, it is
sufficient to carry out the integration involved in $\langle j|\psi
_g \rangle$ only in the domain $\pi /2\leq \theta \leq 3\pi /2$. In
this domain the eigenstates of $H_{evol}$ can be rewritten as
\begin{equation}
 \begin{array}{lll}
 |j\rangle
 &=&(-1)^{(j-1)/2}\sqrt{2/L} \\
 &&\cos [j\pi R(\theta -\pi )/L] \ \ \
(\mbox{if } j \mbox{ is odd})
 \end{array}
 \label{e_Jodd}
\end{equation}
or
\begin{equation}
 |j\rangle =(-1)^{j/2}\sqrt{2/L}\sin [j\pi R(\theta -\pi )/L] \ \ \
(\mbox{if } j \mbox{ is even})
 \label{e_Jeven}
\end{equation}

From Eqs.(\ref{e_Jodd}) and (\ref{e_Jeven}), obviously, $|j\rangle$
is symmetric with respect to $\theta =\pi$ if $j$ is odd, or
antisymmetric if $j$ is even. Since the real (imaginary) part of
$|k\rangle$ is symmetric (antisymmetric) with respect to $\pi$, and
since $\psi _g$ is a superposition of $|k\rangle$, the symmetries
Eqs.(\ref{e_Jodd}) and (\ref{e_Jeven}) lead to a fact that $\langle
j|\psi _g\rangle$ would be a real number if $j$ is odd, or an
imaginary number if $j$ is even. Therefore, when $j\pm j'$ are even,
the product $\langle \psi _{g}|j'\rangle \langle j|\psi _g\rangle$
is a real number, and the time-dependent factor
$e^{-i(j-j')(j+j')\tau }$ in Eq.(\ref{e_Rho}) can be thereby
rewritten as $\cos [(j-j')(j+j')\tau ]$, where $(j-j')(j+j')$ is an
even integer $I_e$. Alternatively, when $j\pm j'$ is odd, the
product $\langle \psi _g|j'\rangle \langle j|\psi _g\rangle $ is an
imaginary number, and the time-dependent factor can be thereby
rewritten as $-\sin [(j-j')(j+j')\tau ]$ where $(j-j')(j+j')$ is an
odd integer $I_o$. Since $\cos [I_e(\pi /2-\tau )]=\cos [I_{e}(\pi
/2+\tau )]$ and $\sin [I_o(\pi /2-\tau )]=\sin [I_{o}(\pi /2+\tau
)]$, we arrive at an important feature of the evolution, namely,
\begin{equation}
 \rho (s,\pi /2-\tau )=\rho (s,\pi /2+\tau )
 \label{e_RhoSysPi2}
\end{equation}
which implies a symmetry of time reflection with respect to $\pi
/2$. Furthermore, since $\sin [p_{j}(L-s)]=(-1)^{j+1}\sin (p_{j}s)$
and $\cos [p_{j}(L-s)]=(-1)^{j}\cos (p_{j}s)$, we have
\begin{equation}
 \rho (L-s,\pi -\tau )=\rho (s,\pi +\tau )
 \label{e_RhoSysPiX}
\end{equation}
which implies a symmetry of time reflection with respect to $\pi$
together with a spatial reflection with respect to $L/2$.

Similarly, one can prove
\begin{eqnarray}
 J(s,\pi /2-\tau )=-J(s,\pi /2+\tau ) \label{e_JSysPi2} \\
 J(L-s,\pi -\tau )=J(s,\pi +\tau )  \label{e_JSysPiX}
\end{eqnarray}
Due to Eqs.(\ref{e_RhoSysPi2}) and (\ref{e_JSysPi2}), the evolution
in the duration $[\pi /2,\pi ]$ is the time-reversal of that in
$[0,\pi /2]$. Due to Eqs.(\ref{e_RhoSysPiX}) and (\ref{e_JSysPiX}),
the evolution in $[\pi ,2\pi ]$ is the time-reversal of that in
$[0,\pi ]$ together with a spatial reflection against $L/2$.
Therefore, the whole evolution can be understood if that in $[0,\pi
/2]$ is clear. In what follows the study is restricted in $[0,\pi
/2]$.

In order to have numerical results, let $m^*=0.063me$ (for InGaAs),
$L=2000nm$, $l=400nm$, $R=1200nm/\pi$, and $V_{o}=30G=0.1243meV$,
$d=\pi/4$. With these parameters, the time unit $t_{o}=4.408\times
10^{-10}\sec$. These choices are rather arbitrary.

\begin{figure}[tbp]
 \centering
 \resizebox{0.95\columnwidth}{!}{\includegraphics{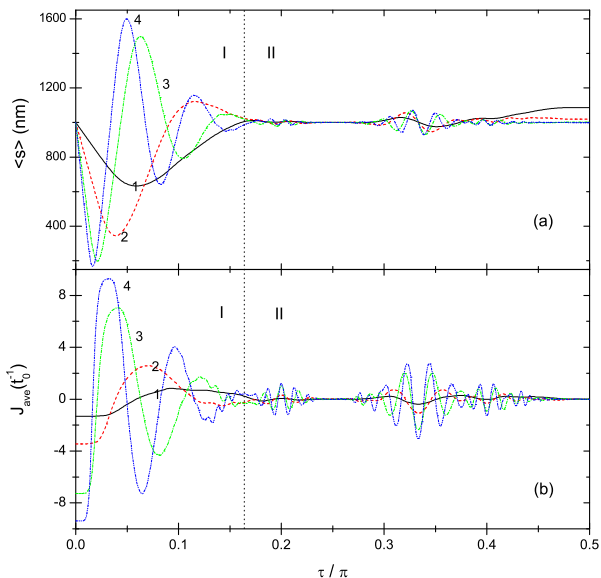}}
 \caption{(Color online.) The time-dependence of the average position
of the electron $\langle s\rangle$ and the average current $J_{ave}$
plotted in a quarter of a period. The curves "1" to "4" have $\Phi
=1.25$, $3.25$, $6.85$, and $8.85$, respectively. }
 \label{eufig2}
\end{figure}
In order to have a general impression on the evolution, we define
the time-dependent average position of the electrons
\begin{equation}
 \langle s\rangle\equiv \int \rho (s,\tau )s\ ds
 \label{e_Save}
\end{equation}
and define the time-dependent average current\ as
\begin{equation}
 J_{ave}(\tau )\equiv \frac{1}{L}\int J(s,\tau )\ ds
 \label{e_Jave}
\end{equation}
When $\Phi$ is given at four values, $\langle s\rangle$ and
$J_{ave}(\tau )$ against $\tau$ are shown in Fig.\ref{eufig2}. When
$\tau =0$, obviously $\langle s\rangle=L/2$ (at the bottom) as shown
in Fig.\ref{eufig2}a. However, the current is not zero initially
($J_{ave}(0)\neq 0$). This is shown in Fig.\ref{eufig2}b, where a
larger $\Phi$ leads to a more negative $J_{ave}(0)$.

Let us define $E_{evol}=\langle\psi _g|H_{evol}|\psi _g\rangle$
which is the energy contained in $\psi _g$ after the transformation
of the Hamiltonian. It was found that, when $\Phi$ increases from
$1.25$ to $8.85$, $E_{evol}$ increases from $0.0152$ to $0.333meV$.
Obviously, a larger $E_{evol}$ will lead to a stronger initial
current which motivates the evolution afterward.

In Fig.\ref{eufig2} the duration $[0,\pi /2]$ can be roughly divided
into two, $[0,\pi /6]$ and $[\pi /6,\pi /2]$. In the former $\langle
s\rangle$ goes toward the two ends alternately. Accordingly,
$J_{ave}$ appears as negative and positive alternately. Obviously,
this implies that the electron oscillates end-to-end repeatedly.
When $\Phi $ is larger, the first minimum of $\langle s\rangle$ will
shift down and left as shown in Fig.\ref{eufig2}a. It implies that,
once the evolution begins, the electron will be closer to A in a
shorter time. Furthermore, a larger $\Phi$ will cause more rounds of
oscillation taking place in the duration $[0,\pi /6]$.

In the duration $[\pi /6,\pi /2]$, $\langle s\rangle$ remains to be
close to $L/2$, and $|J_{ave}|$ becomes small. It implies that the
probabilities of the electron staying at the left and right sides of
the bottom are nearly equal, and both negative and positive currents
might appear in the path simultaneously (this leads to a
cancellation and thereby a smaller $|J_{ave}|$). The above two
durations, for simplicity, are called duration of oscillation (DoS)
and duration of cruise (DoC).

\begin{figure}[tbp]
 \centering
 \resizebox{0.95\columnwidth}{!}{\includegraphics{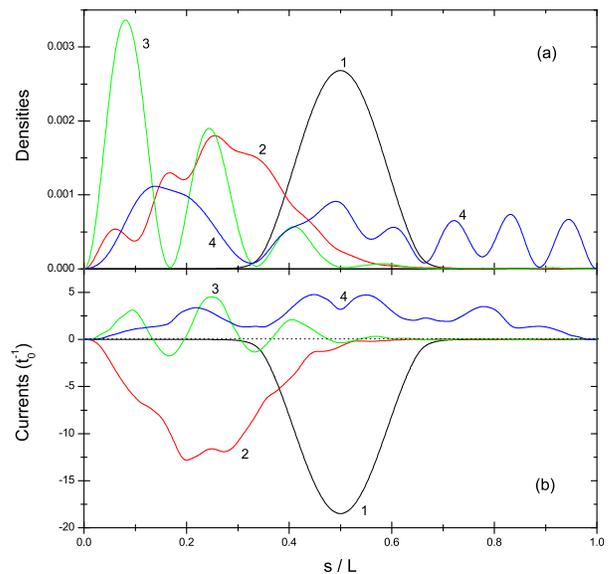}}
 \caption{(Color online.) Spatial distribution of the densities
$\rho (s,\tau )$ and currents $J(s,\tau )$ against $s$. The curves
"1" to "4" have $\tau =0$, $\pi /48$, $\pi /24$, and $\pi /12$,
respectively (these values are within the duration of oscillation
DoS). The preset flux $\Phi =3.25$ (the same for Figs.\ref{eufig3}
to \ref{eufig6}). The unit of current is $t_{o}^{-1}$, where
$t_{o}\equiv 2m^* L^2 /(\pi ^2 \hbar )$ is the unit of time used in
this paper. }
 \label{eufig3}
\end{figure}
To study in detail the evolution in the DoS, the spatial
distribution of $\rho (s,\tau )$ is plotted in Fig.\ref{eufig3},
where $\tau$ is given at four values in the early stage of
evolution. When $\tau=0$, the curve "1" of Fig.\ref{eufig3}a shows
the initial localization, and "1" of Fig.\ref{eufig3}b shows the
strong negative initial current created via the sudden change of the
Hamiltonian. Due to the negative current, the peak of $\rho$ shifts
left rapidly as shown by "2" of Fig.\ref{eufig3}a. Correspondingly,
the distribution of $J$ shifts also left as shown by "2" of
Fig.\ref{eufig3}b. Afterward, the peak of $\rho$ keeps going left
and will be close to A as shown by "3" of Fig.\ref{eufig3}a.
However, "3" of Fig.\ref{eufig3}b is mainly positive. It implies
that the direction of motion has already been reversed and the
electron begins going toward the other end E. A little time later,
the density is partially close to E as shown by "4" of
Fig.\ref{eufig3}a. Meanwhile, the current is positive throughout the
path as shown by "4" of Fig.\ref{eufig3}b. Accordingly, the
distribution of $\rho$ as a whole is going right. Fig.\ref{eufig3}
shows only the first round of oscillation in the DoS when $\Phi
=3.25$.

The maximal initial current $J(L/2,0)$ (e.g., the dip of "1" of
Fig.\ref{eufig3}b) was found to be nearly linearly proportional to
$\Phi $ (with our parameters, $J(L/2,0)\approx -5.69\Phi /t_o$).
Incidentally, if $\Phi <0$, the initial current would be positive
and therefore the direction of motion would be reversed.

\begin{figure}[tbp]
 \centering
 \resizebox{0.95\columnwidth}{!}{\includegraphics{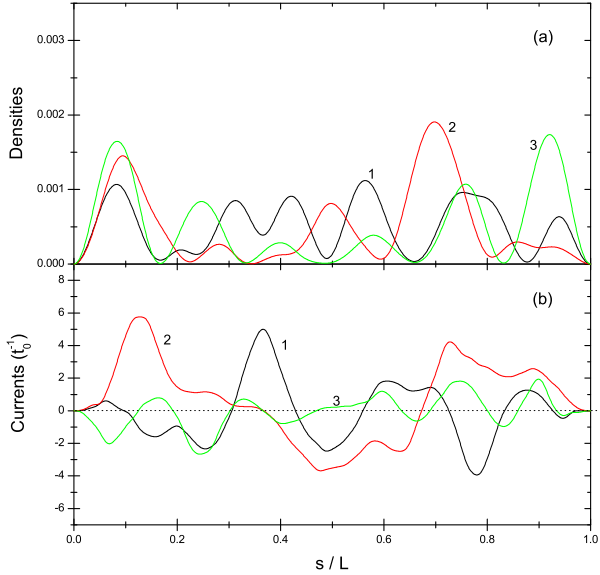}}
 \caption{(Color online.) The same as Fig.\ref{eufig3} but in the
duration of cruise DoC $[\pi /6,\pi /2]$. The curves "1" to "3" have
$\tau =0.285\pi$, $0.354\pi$, and $0.458\pi$, respectively. }
 \label{eufig4}
\end{figure}
In the DoC both $\rho$ and $J$, in general, are widely distributed
along the path from A to E with numbers of peaks and dips as shown
in Fig.\ref{eufig4}. In this duration the current may be positive
(going right) somewhere and negative (going left) elsewhere. If
$J=0$ and $\frac{\partial}{\partial s}J>0$ take place at $s=s_o$,
there would be a source at $s_o$ from where the current flows out to
both sides. Whereas if $J=0$ and $\frac{\partial}{\partial s}J<0$ at
$s_o$, there would be a leak to where the currents flow in from both
sides. Both sources and leaks are found in the path. The classical
picture of motion of the electron in the DoC is not clear, it seems
to be chaotic.

\begin{figure}[tbp]
 \centering
 \resizebox{0.95\columnwidth}{!}{\includegraphics{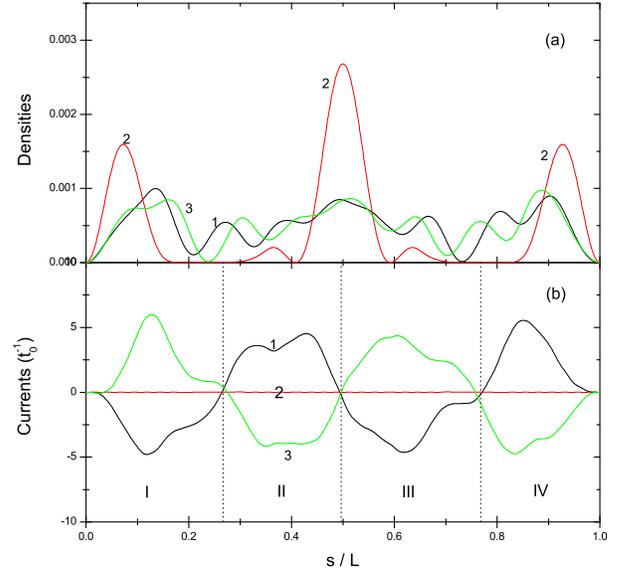}}
 \caption{(Color online.) The same as Fig.\ref{eufig3} but the
curves "1" to "3" have $\tau =\pi /4-\pi /96$, $\pi /4$, and $\pi
/4+\pi /96$, respectively. }
 \label{eufig5}
\end{figure}
However, in the DoC, there is a noticeable instant. When $\tau =\pi
/4$, we have $\langle s\rangle=L/2$ and $J_{ave}=0$ disregarding how
$\Phi$ is as shown in Fig.\ref{eufig2}. At this instant $\rho$ is no
more widely distributed but concentrated close to the two ends and
the bottom C as shown by "2" of Fig.\ref{eufig5}a (where the peak at
the bottom is much higher). When $\tau$ is a little earlier and
later than $\pi /4$, $\rho$ and $J$ are shown, respectively, by "1"
and "3" of Fig.\ref{eufig5}a and \ref{eufig5}b. They together
demonstrate how the density is concentrated into three peaks and
spread out afterward. The abscissa of Fig.\ref{eufig5}b has been
divided into four regions. The curve "1" of Fig.\ref{eufig5}b is
negative in region I. Thus the density is pushed left resulting in
forming the peak of $\rho$ close to A. Besides, "1" is positive in
II but negative in III. Thus $\rho$ is pushed from both sides of C
to the bottom resulting in forming the highest peak. Furthermore,
the positive current in IV leads to the peak\ close to E. Exactly at
the instant $\tau =\pi /4$, the current is zero as shown by "2" of
Fig.\ref{eufig5}b. It implies that the system is static
instantaneously. Afterward, the current increases rapidly but in
reverse direction as shown by "3" of Fig.\ref{eufig5}b. This leads
to the disappearance of the three peaks as shown by "3" of
Fig.\ref{eufig5}a. The existence of instants wherein the system is
instantaneously static is not at all surprising, this might happen
in various types of oscillation (say, in classical pendulum).

\begin{figure}[tbp]
 \centering
 \resizebox{0.95\columnwidth}{!}{\includegraphics{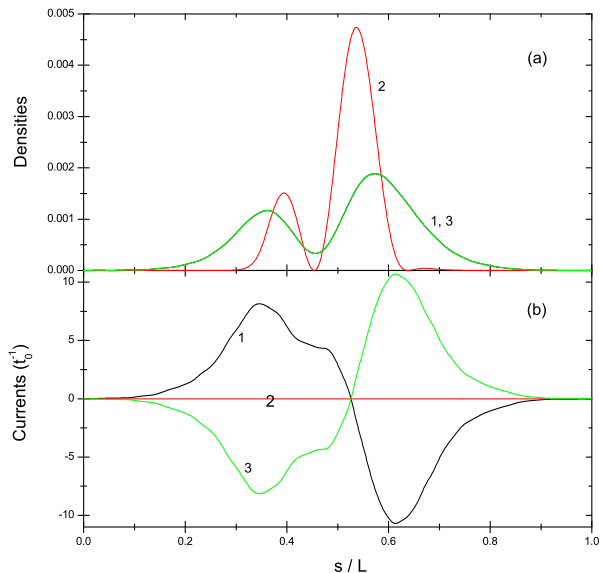}}
 \caption{(Color online.) The same as Fig.\ref{eufig3} but the
curves "1" to "3" have $\tau =\pi /2-\pi /96$, $\pi /2$, and $\pi
/2+\pi /96$, respectively. }
 \label{eufig6}
\end{figure}
When $\tau$ is close to $\pi /2$, the density is once again
concentrated in the bottom as shown by "2" of Fig.\ref{eufig6}a,
where the main peak is very sharp and high. At the same time, the
associated current ("2" of Fig.\ref{eufig6}b) is exactly zero (this
is obvious from Eq.(\ref{e_JSysPi2})). Thus the system becomes once
again static instantaneously. "1" and "3" of 6a describe the
densities a little earlier and later, respectively, than $\pi /2$.
They overlap exactly due to the symmetry given in
Eq.(\ref{e_RhoSysPi2}). "1" and "3" of Fig.\ref{eufig6}b are exactly
opposite to each other due to Eq.(\ref{e_JSysPi2}). In fact, they
cause a rapid gathering and, successively, a rapid extension of
$\rho$. In particular, the exact symmetry appearing in
Fig.\ref{eufig6}b implies that \textit{the evolution has arrives at
a turning point. Afterward, the evolution will proceed exactly
reversely as demonstrated by Eqs.(\ref{e_RhoSysPi2}) and
(\ref{e_JSysPi2})}. For examples, the density will be re-gathered
into three peaks exactly as "2" of Fig.\ref{eufig5}a when $\tau
=3\pi /4$, and will re-visit A exactly as "3" of 3a when $\tau =\pi
-\pi /24$, etc. When $\tau =\pi $, $\rho$ is exactly as "1" of 3a,
while $J$ is exactly opposite to "1" of Fig.\ref{eufig3}b. Thus,
instead of going left, the peak of $\rho$ at the bottom goes right
at $\tau =\pi$ due to the positive $J$.

It is noted that from Eqs.(\ref{e_RhoSysPi2}) and
(\ref{e_RhoSysPiX}), we have
\begin{equation}
 \rho (L-s,\tau )=\rho (s,\pi +\tau )
 \label{e_RhoLtau}
\end{equation}
From (\ref{e_JSysPi2}) and (\ref{e_JSysPiX}), we have
\begin{equation}
 J(L-s,\tau )=-J(s,\pi +\tau )
 \label{e_JLtau}
\end{equation}
In the duration $[\pi ,2\pi ]$, Eqs.(\ref{e_RhoLtau}) and
(\ref{e_JLtau}) together imply that the evolution is just a repeat
of that from $0$ to $\pi$, but with an interchange
$s\longleftrightarrow L-s$ (namely, an interchange of the A and E
ends). Thus, the evolution in the whole period $2\pi$ is completely
clear, and it goes on again and again periodically.

Now let us compare the current in the U-oscillator with the famous
Aharonov-Bohm (A-B) persistent current of an electron on a ring. The
maximal A-B current of the ground state is $J_{A-B}=\hbar /(4\pi m^*
R^2)$. Thus, $t_{o}^{-1}=2\pi (L-2l)^2 /L^2 \ J_{A-B}$. For our
parameters, we found that $J(L/2,0)=-12.87\Phi J_{A-B}$. Thus, giant
current can be obtained if $\Phi$ is sufficiently strong (say, if
$\Phi \approx 10$, $J(L/2,0)$ is two orders stronger than the A-B
current).

\begin{figure}[tbp]
 \centering
 \resizebox{0.95\columnwidth}{!}{\includegraphics{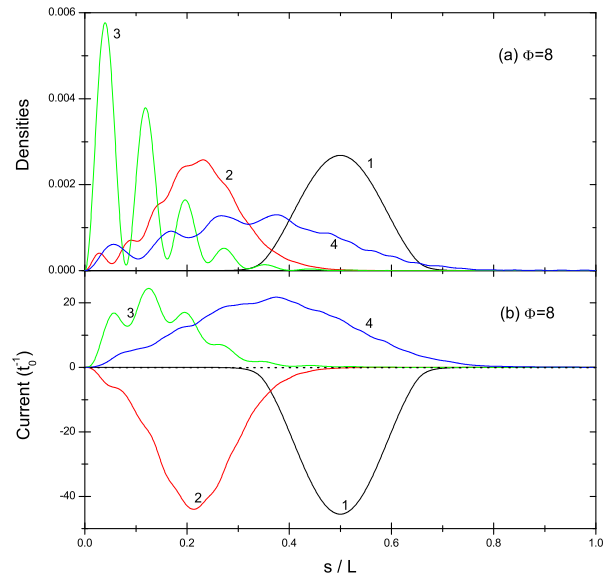}}
 \caption{(Color online.) $\rho (s,\tau )$ and $J(s,\tau )$ with
$\Phi =8$. The curves "1" to "4" have $\tau =0$, $\pi /96$, $\pi
/48$, and $\pi /32$, respectively. }
 \label{eufig7}
\end{figure}
Since $t_o$ is proportional to $m^* L^2$, a longer path and/or a
heavier effective mass will lead to a longer period. The preset
$\Phi$ as a motivity is crucial to the oscillator. Its effect has
been shown in Fig.\ref{eufig2} and is further shown in
Fig.\ref{eufig7} with $\Phi =8$ (to be compared with Fig.3 with
$\Phi =3.25$). Comparing the distributions of "3" of
Fig.\ref{eufig7}a and Fig.\ref{eufig3}a, the former is closer to A
and occurs much earlier (the electron rushes to A more rapidly).
While $\Phi$ is a very sensitive factor, the parameters of the
preset potential is less sensitive to the evolution.

\begin{figure}[tbp]
 \centering
 \resizebox{0.95\columnwidth}{!}{\includegraphics{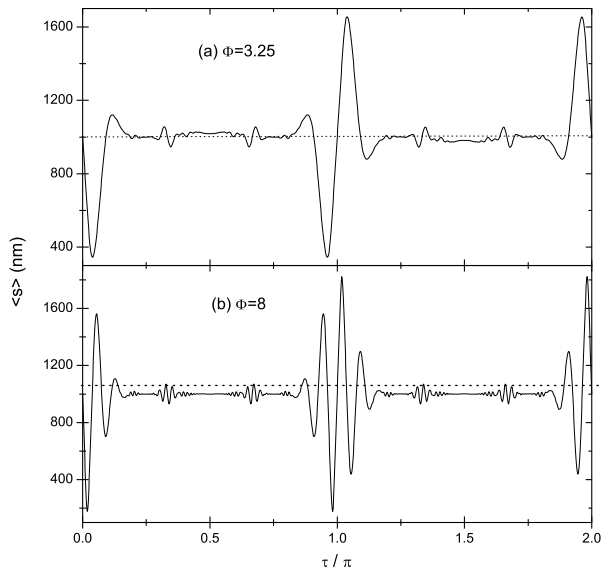}}
 \caption{The time-dependence of $\langle s\rangle$ in a whole
period. $\Phi =3.25$ (a) and $8$ (b). }
 \label{eufig8}
\end{figure}
In summary, a U-oscillator has been proposed and studied. The motion
of electron in the device is strictly periodic and adjustable. The
following points are reminded

(i) When $t_o$ is used as the unit of time, the period is $2\pi$.
The evolution in the second quarter of a period  is the
time-reversal of that of the first quarter, and the evolution in the
second half is the same as the first half but with the interchange
$s\longleftrightarrow L-s$, namely, a spatial reflection against
$L/2$.

(ii) Each quarter can be divided into two durations, namely, DoS and
DoC. The classical picture in the former is clear (a few rounds of
oscillation appear), but not clear in the latter. When $\Phi$ is
larger, more rounds of oscillation will appear in the DoS.

(iii) Let the time associated with the lowest minimum of Fig.2a be
$\tau _{end}$. Then, in the first half period, the electron is very
close to A twice at $\tau _{end}$ and $\pi -\tau _{end}$,
respectively. Whereas in the second half period, the electron is
very close to E also twice at $\pi +\tau _{end}$ and $2\pi -\tau
_{end}$, respectively. Such a close contact of the electron with the
two ends in a whole period is shown in Fig.\ref{eufig8}, where the
appearance of rounds of oscillation in the DoS is also clear. On the
contrary with classical pendulum, it is clear from Fig.\ref{eufig8}
that is explicitly the end-to-end oscillation occurs only if $\tau$
is close to $I\pi$ ($I$ is a positive integer).

(iv) Each time when $\tau =I\pi /4$ , the system is instantaneously
static, namely, $J(s,I\pi /4)=0$ for all $s$. Meanwhile, the density
would be mostly concentrated in the neighborhood of C, namely, the
bottom of the U-pipe.

(v) The introduction and the sudden removal of $\Phi$ is a crucial
point. This leads to the sudden creation of the initial current,
which motivates the evolution afterward. In particular, when $\Phi$
is sufficiently large, giant periodic current (two or more orders
stronger than the A-B current) can be obtained.

The device might work as a quantum pendulum. It can be generalized
to include a group of localized electrons initially. This case
deserves to be further studied. The idea proposed in this paper
might be useful in micro technology.

\begin{acknowledgments}
Acknowledgment: The support from NSFC under the grant 10574163 and
10874249 is appreciated.
\end{acknowledgments}

\end{document}